ASTRONOMY

# *Improving Planet-finding Spectrometers*

Like the miniaturization of modern computers, next-generation radial velocity instruments will be significantly smaller and more powerful than their predecessors

*By* **Justin R. Crepp**[1]

Adaptive optics (AO) systems correct for optical wave-front errors introduced by Earth's turbulent atmosphere, turning initially blurry images into intense diffraction-limited concentrations of light. Over the past decade, the implementation of AO systems on the world's largest telescopes has revolutionized essentially all areas of astronomy *(1)*. Instruments that receive a well-corrected beam of light can operate as if observing from space and thus benefit from an order of magnitude higher spatial and spectral resolution (Fig. 1).

Given the benefits of working with non-fuzzy images, it may therefore be surprising to learn that one of the most important and venerable techniques for finding extrasolar planets – the Doppler radial velocity method – still uses "seeing-limited" observations, i.e., measurements obtained without AO correction. Doppler instruments began detecting extrasolar planets in the mid-1990's, well before AO systems became commonplace at astronomical facilities. Thus radial velocity measurements were by default seeing-limited *(2)*. The targeting of Sun-like stars in search of solar-system analogues and development of the iodine gas cell for velocity calibration (which only operates in the $\lambda=0.50-0.62$ μm wavelength range) further reinforced any initial "Copernican-esque" tendencies to observe at visible wavelengths, thus precluding diffraction-limited capabilities from the ground.

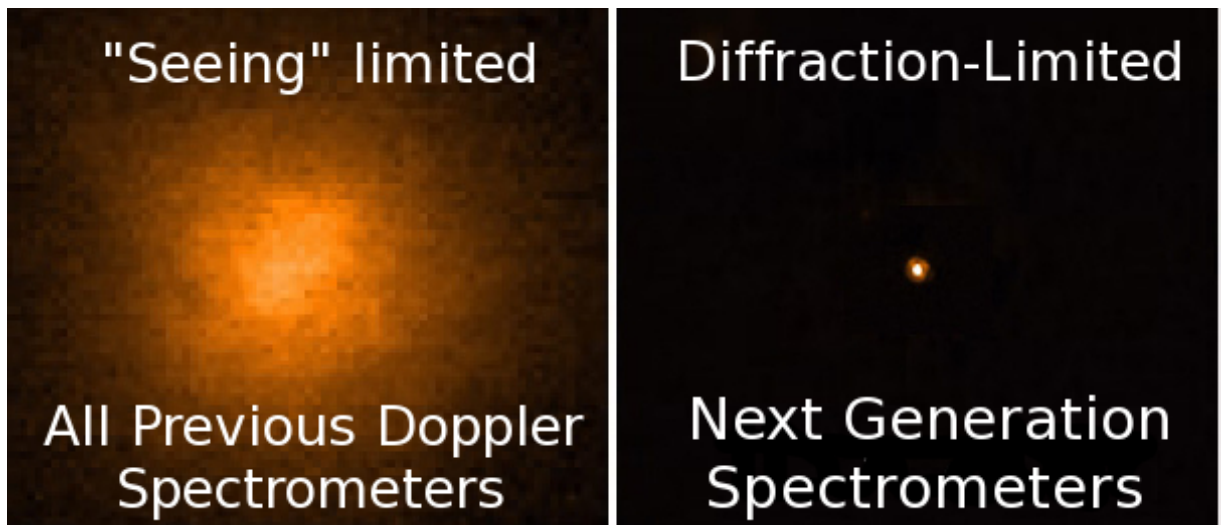

**Seeing-limited images versus diffraction-limited images.** An AO-fed spectrometer achieves much higher spatial resolution, and hence spectral resolution, fundamentally changing its design compared to presently available Doppler instruments. All such design modifications lead to improved velocity precision. Image adapted from Takami 2008.

[1] Department of Physics, University of Notre Dame, Notre Dame, IN 46556, USA
Email: jcrepp@nd.edu

The Doppler method has since matured and observations have naturally branched out to study planets orbiting different types of stars *(3)*. For example, we now know that the lowest-mass stars in the galaxy, the M-dwarfs, have a propensity to form terrestrial planets *(4)*. NASA's Kepler spacecraft has corroborated this result finding that many of these worlds orbit near the habitable zone *(5)*. As such, several teams are now developing the first precision infrared spectrometers to gain access to the black body radiation of M-dwarfs which peaks in the $\lambda$=1-2 µm range *(6, 7)*.

While the scientific motivation for building radial velocity instruments in the infrared is clear, there exists an important threshold-crossing that occurs when observing at wavelengths just beyond $\lambda = 1.0$ µm: the ability to achieve diffraction-limited performance by correcting for atmospheric turbulence using AO. Instruments designed to operate behind an AO system have fundamentally different properties than their seeing-limited counterparts. In particular, their physical size shrinks commensurately with the size of the input image. It turns out that all such hardware modifications resulting from this effect help to improve velocity precision.

For example, one of the most problematic sources of error encountered by present-day Doppler instruments involves their stability: precise velocity time-series measurements require a spectrometer with exquisite temperature and pressure control. Any changes within the instrument optical path – tiny movements of lenses, mirrors, gratings, filters, etc. – can mimic a Doppler shift. To detect the signal of an Earth-like planet orbiting in the habitable zone, thermal fluctuations must be reduced to the milli-Kelvin level *(8)*.

The size of a seeing-limited spectrometer grows with telescope diameter. For 8-10m class facilities, it is not uncommon for such instruments to be the size of a small car, making precise environmental control challenging. However, a variety of creative control solutions exist to overcome this challenge for spectrometers operating at the diffraction-limit. For instance, instruments that are an order of magnitude smaller in all directions can be given significant "thermal inertia" by enclosing them in a bath of liquid cryogens. Also, more exotic materials may be used like Invar – which derives from the word "invariable", an iron-nickel alloy having a coefficient of thermal expansion 10 times lower than the Aluminum-based materials normally used for astronomical instruments. Further, a compact optical design is less expensive, benefits from components with higher optical quality, has a faster development timescale, and can achieve an ultra-high vacuum ($P < 10^{-8}$ Torr).

Diffraction-limited spectrometers also solve a host of other problems in addition to stability. Background contamination from the Moon, the sky, and unseen neighboring stars is reduced by three orders of magnitude. "Modal noise" from fiber optic cables, introduced by changes in the interference pattern of starlight, can be eliminated entirely by virtue of using single mode fibers. And, by working in the infrared, "astrophysical jitter" caused by star spots that rotate in and out of view as a star spins is reduced by a factor of several compared to visible light instruments. These systematic effects dominate the error budgets of present-day radial velocity spectrometers.

The singular challenge to realizing a diffraction-limited Doppler instrument involves coupling starlight into small (single mode) optical fibers. This barrier has recently been overcome by Jovanovic et al. 2014 *(9)*. Using a new "extreme" AO system built for the Subaru 8.2m telescope, single mode fiber coupling efficiencies of 68% have been demonstrated with an 8 μm diameter fiber core, approaching the theoretical maximum attainable limit set by the level of delivered wave-front correction. The observations were acquired in broadband light for wavelengths near λ=1.55 μm, just longwards of the ideal λ=1.0-1.1 μm window (the Y-band) where the Earth's atmosphere is uniformly transmissive and not riddled with OH-emission lines *(10)*. The implications of these measurements are that Doppler spectrometers the size of a mailbox are feasible and that instrument throughput will be sufficient to permit record-breaking precision using existing telescope facilities.

Indeed, a substantive improvement in velocity precision would be timely. NASA's forthcoming Transiting Exoplanet Survey Satellite (TESS) will survey the entire night sky, revealing the addresses of several hundred nearby rocky planets that may resemble Earth *(11)*. The majority of the terrestrial worlds found to reside near the habitable zone will induce Doppler reflex motions well below the detection limits of presently available spectrometers.

Next-generation Doppler instruments promise to be significantly smaller and more powerful than their seeing-limited counterparts. Despite being optimized for near-infrared wavelengths, diffraction-limited spectrometers will not only be ideally suited for M-stars, but can also study nearby Sun-like stars with unprecedented detail. The errors that conspire to overwhelm the Doppler signal of an Earth-analogue can be mitigated by developing instruments that benefit from the unimpaired vision of telescopes that correct for atmospheric turbulence using AO.

**REFERENCES**
1. R. Davies, M. Kaspar, *Annu. Rev. Astron. Astrophys.* **50**, 305 (2012)
2. M. Mayor, D. Queloz, *Nature* **378**, 355 (1995)
3. J. Johnson et al., *PASP* **122**, 905 (2010)
4. X. Bonfils et al., *AA* **549**, 109 (2013)
5. J. Johnson, *Physics Today* **67**, 31 (2014)
6. A. Quirrenbach et al., *Proc. SPIE*, **8446**, 13 (2012)
7. S. Mahadevan et al., *Proc. SPIE*, **8446**, 14 (2012)
8. C. Lovis, D. Fischer, *Exoplanets Book*, Ed. Seager, **526**, 27 (2011)
9. N. Jovanovic et al., *Proc. SPIE*, **9147**, 287 (2014)
10. S. Redman, PhD Thesis, ISBN: 9781303582875 (2011)
11. G. Ricker et al., *EBI Conf. Proc.*, arXiv:1406:0151 (2014)
12. H. Takami, *Proc. SPIE*, **7014**, 04 (2008)